\documentclass[preprint, 12pt]{elsarticle}

\usepackage{amssymb}
\usepackage{amsmath}
\usepackage{mathtools}
\usepackage{siunitx}

\usepackage[margin=1in]{geometry}

\usepackage{lineno}

\usepackage{multirow}%
\usepackage{mathrsfs}%
\usepackage{appendix}%
\usepackage{xcolor}%
\usepackage{textcomp}%
\usepackage{manyfoot}%
\usepackage{booktabs}%
\usepackage{algorithm}%
\usepackage{algorithmic}%
\usepackage{listings}%
\usepackage{url}

\usepackage{subcaption}
\usepackage[capitalize]{cleveref}
\usepackage{bm}
\usepackage{xfrac}
\usepackage{parskip}
\usepackage{colortbl}
\usepackage{placeins}

\newcommand{\adaptedparams}{\params^{\prime}}
\newcommand{\metagrad}{\nabla_{\initparams}\mathcal{L}_{\text{meta}}}

\newcommand{\params}{\boldsymbol{\theta}}

\newcommand{\initparams}{\params_{0}}
\newcommand{\density}{\boldsymbol{\rho}} \newcommand{\x}{\mathbf{x}}

\newcommand{\task}{\mathcal{T}_{i}}
\DeclareMathOperator*{\argmin}{\arg\!\min\;}

\journal{Engineering Applications of Artificial Intelligence}

\begin{document}

\begin{frontmatter}
    \title{Meta-neural Topology Optimization: Knowledge Infusion with
        Meta-learning} 

    \author[a,b]{Igor Kuszczak}
    \author[c]{Gawe\l{} Ku\'{s}}
    \author[a,b]{Federico Bosi\corref{cor1}}
    \ead{f.bosi@ucl.ac.uk}
    \author[c]{Miguel A. Bessa\corref{cor1}}
    \ead{miguel_bessa@brown.edu}

    \cortext[cor1]{Corresponding author}

    \affiliation[a]{organization={Department of Innovative Technologies,
            University of Applied Sciences and Arts of Southern Switzerland},
        addressline={Via la Santa 1}, city={Lugano}, postcode={6962},
        country={Switzerland}}
    \affiliation[b]{organization={Department of Mechanical Engineering,
            University College London}, addressline={Torrington Place},
        city={London}, postcode={WC1E 7JE}, country={United Kingdom}}
    \affiliation[c]{organization={School of Engineering, Brown University},
        addressline={184 Hope Street}, city={Providence}, postcode={RI 02912},
        country={United States}}

    \begin{abstract}
        When faced with novel design problems, traditional topology optimization
        methods discard all prior design experience and start from a uniform
        initial guess. While this avoids biasing the optimizer towards any
        particular solution, it also means that many computationally expensive
        iterations are needed to converge. Existing machine learning approaches
        address this through data-driven design prediction, but require large
        datasets of pre-optimized structures and often struggle to generalize
        across boundary conditions and mesh resolutions. We propose a new
        method, termed meta-neural topology optimization, which uses a
        meta-learning algorithm to learn effective initial designs for topology
        optimization with neural field parameterizations --- continuous,
        mesh-independent representations that encode material distributions in
        the weights of a neural network. Through bilevel optimization, our
        method distills reusable design knowledge from partial optimization
        trajectories, eliminating the need for pre-optimized training data. By
        conditioning the neural field on strain energy fields of reference
        designs, a single set of learned parameters encodes problem-specific
        initial structures for diverse boundary conditions. We evaluate our
        approach on 3000 compliance minimization tasks across in-distribution,
        out-of-distribution, and cross-resolution scenarios. Our method
        converges in fewer iterations in 57.6\% of in-distribution and 74.1\% of
        cross-resolution tasks, while maintaining design quality competitive
        with standard density-based topology optimization. Notably,
        initializations learned on coarse meshes transfer successfully to
        discretizations four times finer. Code is available at
        \url{https://github.com/bessagroup/metatopia}.
    \end{abstract}

    \begin{graphicalabstract}
        \includegraphics[width=\linewidth]{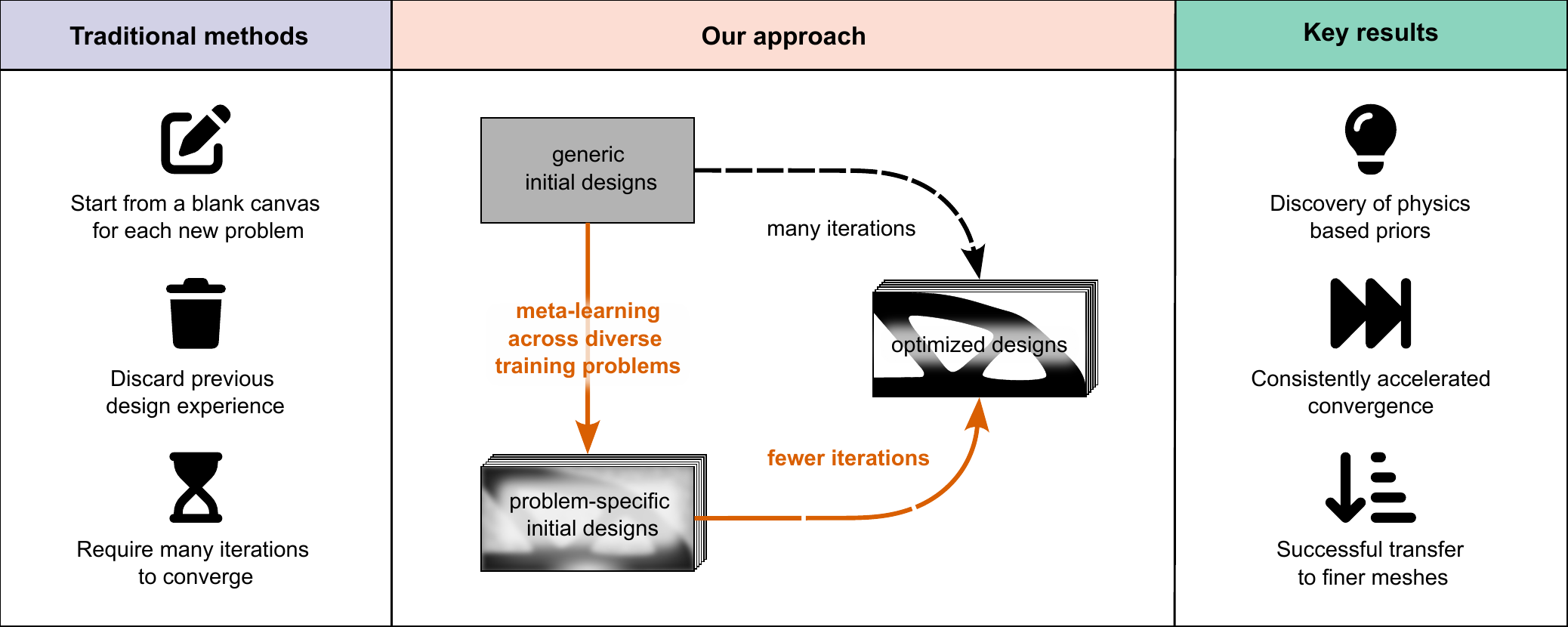}

    \end{graphicalabstract}

    \begin{highlights}
        \item Meta-learning infuses past optimization experience into topology
        optimization.
        \item Learned initializations consistently accelerate neural topology
        optimization.
        \item This narrows the gap between standard and neural topology
        optimization.
        \item Initializations learned on coarse meshes transfer to unseen finer
        meshes.
    \end{highlights}


\end{frontmatter}



\section{Introduction}\label{introduction} Topology optimization (TO)
has transformed computational design across multiple scales and
disciplines. Its successful applications span from nanometer-scale
optical cavities that transcend traditional limitations on light
confinement~\citep{albrechtsen2022} to full-scale aircraft wing designs
with unprecedented structural efficiency~\citep{aage2017}. While the
field has seen significant advancements since the pioneering work
by~\citet{bendsoe1988}, one fundamental aspect of human design
capability remains underexplored: learning from experience.

Traditional TO approaches adopt a \textit{tabula rasa} strategy, reinitializing the design
variables before every run. This practice persists despite substantial
evidence that strategic initial guesses can accelerate
convergence~\citep{sigmund2013}. Since each TO iteration requires at
least one computationally expensive finite element analysis, and many
iterations are needed to reach optimized designs, the lack of knowledge
reuse represents a significant missed opportunity.

To develop TO methods that learn from experience, we must first
understand the limitations of knowledge transfer in standard approaches.
In density-based methods~\citep{bendsoe1989}, the design domain is
discretized into finite elements, and each element is assigned a density
value between 0 and 1, indicating the material occupancy. The density
distribution is then optimized for the selected objective using a
gradient-based optimizer, commonly the Method of Moving Asymptotes
(MMA)~\citep{svanberg1987}. Since the design variables are defined on
the computational mesh, they must be redefined for each new geometry and
mesh resolution~\citep{white2020}, preventing the direct transfer of
learned representations.

Neural topology optimization addresses this limitation by encoding the
density distribution in the learnable parameters of a neural network.
Approaches based on neural fields~\citep{xie2022} provide continuous,
mesh-agnostic representations by mapping spatial coordinates $\x$ to
density values $\density\left(\x;\, \params \right) = f_{\params}(\x)$,
enabling parameters to transfer across different geometries and
resolutions. Neural TO has further been shown to reshape the optimization
landscape, uncovering designs that mesh-based parameterizations cannot
access~\citep{herrmann2024}. However, this reshaping cuts both ways:
neural parameterizations can create highly non-convex landscapes that
significantly hinder convergence, even in simple compliance minimization
problems~\citep{sanu2024}, rendering the quality of the initial
parameters critical.

Together, these properties create both the opportunity and the need for
learned initializations in neural TO. Prior neural TO work has relied on
heuristic initializations such as a uniform-density
solution~\citep{zhang2021, doosti2021} or transfer learning from a small
set of pre-solved reference designs~\citep{herrmann2024}. Unlike
transfer learning, meta-learning is designed to distill generalizable
knowledge from a distribution of related tasks~\citep{hospedales2022},
directly optimizing the learning process to accelerate adaptation to
unseen problems. Gradient-based methods such as MAML~\citep{finn2017}
and Reptile~\citep{nichol2018} have been applied to neural fields across
multiple modalities~\citep{sitzmann2020, tancik2021, tack2023} and
engineering problems~\citep{huang2022, penwarden2023, qin2022}, but
meta-learning has not been explored in the context of topology
optimization.

In this work, we introduce meta-neural topology optimization, which
applies meta-learning to neural TO to distill a
reusable network initialization from a distribution of related
design tasks~\citep{hospedales2022}. As surveyed in
Section~\ref{sec:related_work}, our approach is fundamentally different
from existing machine-learning-based TO methods, which require
precomputed datasets of optimized designs. Our contributions are:
\begin{itemize}
    \item \textbf{Meta-learning for neural TO.} We present the first
          application of meta-learning to neural topology optimization,
          learning a reusable initialization from a distribution of related
          tasks with no precomputed designs.
    \item \textbf{Faster convergence and cross-resolution transfer.}
          Meta-learned initializations consistently reduce iterations to
          convergence and transfer from lower- to higher-resolution problems.
    \item \textbf{Emergent physical prior.} Our approach rediscovers
          strain-energy initialization as a physically meaningful prior that
          also accelerates the standard density-based optimizer.
\end{itemize}

\section{Related Work}\label{sec:related_work}
Most machine-learning approaches to topology optimization fall into two
categories: direct-design methods that predict optimized structures in a
single forward pass, and acceleration methods that skip portions of a
classical optimizer. Direct-design approaches span CNN prediction from
coarse-resolution inputs~\citep{yu2019}, conditional GANs with
physical-field conditioning~\citep{nie2021}, neural-field-based approaches
with post-processing via brief SIMP optimization~\citep{nobari2024}, and
latent space diffusion~\citep{lutheran2025, nobari2025}; however, the
predicted topology cannot alter which local optimum the downstream
physics solver reaches. Acceleration methods take various forms,
including predicting final designs from partial SIMP
runs~\citep{sosnovik2017}, mapping early-iteration density histories to
near-converged states~\citep{kallioras2020}, and upsampling coarse-mesh
solutions via super-resolution~\citep{lim2024}. Both categories achieve fast design
synthesis at inference time and, in recent instances, near-SIMP
compliance quality within their training distributions~\citep{nobari2024}.
Despite their methodological diversity, every approach requires a
precomputed dataset of optimized designs generated by a reference
optimizer, inheriting its biases and local-minima structure. The surveys
by \citet{woldseth2022} and \citet{shin2023} provide comprehensive
coverage of both categories.

A separate line of work uses neural networks not to predict designs from
a training set, but as the representation of the density field
itself~\citep{hoyer2019,
chandrasekhar2022a, doosti2021, zhang2021}. Network weights are the
design variables, optimized directly against the physical objective, so
the full topology optimization pipeline is retained and no precomputed
training data is required. This approach offers continuous,
mesh-agnostic representations and freedom from checkerboard
artifacts~\citep{dupuis2021, chandrasekhar2022a}, but also reshapes the
optimization landscape in ways that are not always beneficial.
Initialization is
particularly consequential in this setting: \citet{sanu2024} showed that
neural parameterizations can produce highly non-convex compliance
landscapes, while \citet{herrmann2024} showed that a well-chosen
initialization can reach local optima inaccessible to standard methods.
Specifically, \citet{herrmann2024} pretrained a neural parameterization
based on the U-Net architecture on a small collection of designs
conventionally optimized across a range of acoustic frequencies, and
found that this transfer-learned initialization consistently outperforms
random initialization on acoustic TO. While the approach was effective,
it was limited to a single set of boundary conditions, and still relied
on a small dataset of conventionally-optimized structures.

\section{Methods}
\subsection{Topology optimization} \label{sec:formulation} We focus on
compliance minimization, where the goal is to distribute the material
within a fixed design domain such that the global stiffness of the
resulting structure is maximized. We consider a typical formulation,
where the problem is discretized with finite
elements~\citep{sigmund2013}:

\begin{align}
     & \mathop{\text{minimize}}\limits_{\density} &  & c(\density) = \mathbf{f}^{\intercal}\mathbf{u}(\density)\label{eq:objective} \\[0.5em] \
     & \text{subject to }                         &  & \mathbf{K}(\density)\mathbf{u} = \mathbf{f} \label{eq:equilibrium}           \\
     &                                            &  & \sum_{e=1}^{N} \rho_e v_e \leq V^{*} \label{eq:volume}                       \\
     &                                            &  & 0 \leq \rho_e \leq 1; \quad \forall e \in \{1,\ldots,N\}. \label{eq:box}
\end{align}

Here, $\density$ is a vector of elementwise-constant densities
$\rho_{e}$. The stiffness matrix, global displacement vector, and force
vector are denoted by $\mathbf{K}$, $\mathbf{u}$, and $\mathbf{f}$,
respectively. The target volume is represented by $V^{*}$ with $v_{e}$
denoting the volumes of individual elements $e$.

The global stiffness matrix $\mathbf{K}(\density)$ is assembled from
element stiffness matrices, each scaled by the corresponding elementwise
stiffness $E_{e}$. These are related to densities through the modified
Solid Isotropic Material with Penalization
interpolation~\citep{sigmund2013}:

\begin{equation}
    E_{e}(\rho_{e}) = E_{\text{min}} + \rho_{e}^{p}(E_{\text{max}}-E_{\text{min}}),
\end{equation}

where $E_{\text{max}}$ is the Young's modulus of the base material,
$E_{\text{min}}$ is an artificial stiffness value assigned to the void
regions, and $p$ is a scalar penalization factor. In all our
experiments, we use the standard values of $E_{\text{max}} = 1.0$,
$E_{\text{min}} = 10^{-9}$, $p=3$, and a Poisson's ratio of $\nu = 0.3$.

Optimization follows the nested analysis and design approach, where we
iteratively (i) solve the discretized equilibrium equation
\eqref{eq:equilibrium} for the displacement field $\mathbf{u}$, given a
density distribution $\density$, (ii) evaluate the objective
function~\eqref{eq:objective}, and (iii) update the densities
considering the remaining constraints. In this approach, displacements
$\mathbf{u}(\density)$ are implicit functions of the densities,
satisfying equilibrium at each optimization iteration.

To compute the required gradients, standard topology optimization
implementations derive problem-specific adjoint equations for each
objective function. In our implementation, we instead apply implicit
differentiation to the linear solver~\citep{blondel2022}, which allows
us to compute numerically exact gradients for arbitrary objective
functions while natively supporting higher-order derivatives.

\subsection{Neural TO} \label{sec:neural_to} 
\begin{figure*}[!thb]
    \centering
    \includegraphics[width=120mm]{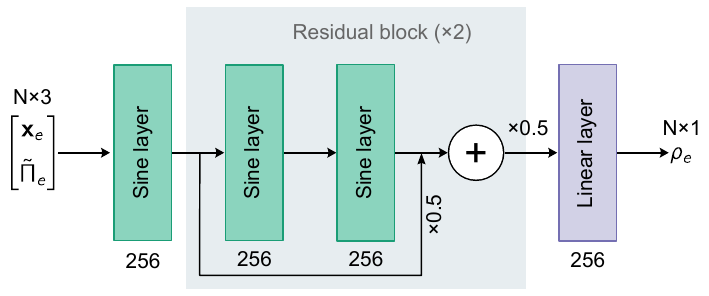}
    \caption{Residual SIREN architecture. The network maps element centroid
        coordinates $\x_{e}$ and strain energy density $\tilde{\Pi}_{e}$ to
        element densities $\rho_{e}$.}
    \label{fig:ressiren}
\end{figure*}

Rather than treating densities as elementwise design variables ---
the direct, mesh-coupled parameterization of standard TO --- we can
conceptualize the material distribution as a continuous field
$\density(\x)$ defined over the design domain $\x \in \Omega$,
independent of the finite element discretization. This continuous
perspective aligns naturally with the physics of the problem: material
exists in space continuously, and the discrete finite element mesh is
merely a computational tool required for the analysis. Neural fields
provide an alternative parameterization, using neural networks to define
smooth functions of spatial coordinates that can be evaluated at
arbitrary locations~\citep{xie2022}. In line with this approach, the density field
can be expressed as
\begin{equation}
    \density(\x, \mathbf{z};\, \params) = f_{\params}(\x,\,\mathbf{z}(\x))
\end{equation}
where $f_{\params}$ is a neural network with learnable parameters
$\params$, and $\mathbf{z}(\x)$ is a conditioning input carrying
problem-specific information. In this work, we reparameterized the
density fields using sinusoidal representation
networks~(SIRENs)~\citep{sitzmann2020} with ResNet-like skip connections
inspired by the Neural Implicit Flow framework~\citep{pan2023}. SIREN
uses periodic activation functions controlled by a single hyperparameter
$\omega_0$ to represent fine structural features, overcoming the
spectral bias observed in standard multilayer perceptrons with ReLU
activations~\citep{rahaman2019}. Our architecture consisted of an input
sine layer followed by two residual blocks of two sine layers each ---
five hidden layers of width 256 in total --- and a linear output
layer~(as shown in~\cref{fig:ressiren}), yielding a total of
$\num{264449}$ learnable parameters. The network size was
selected based on preliminary experiments to balance expressivity and
computational cost. We found that the skip connections preserved the
conditioning signal across network depth and stabilized the
meta-training process.

Following~\citet{chen2023b}, the conditioning input $\mathbf{z}(\x)$ in
our formulation corresponds to the normalized strain energy density of a
reference design $\boldsymbol{\tilde{\Pi}}(\x)$. Here, the reference
design is a uniform density distribution with values corresponding to
the target volume fraction~$V^{*}$. Conditioning enables a single set of
learned parameters $\params$ to represent different structures depending
on the problem specification. For each problem, we solved the
equilibrium equation~\eqref{eq:equilibrium} for a reference design, and
computed the elementwise strain energy densities as:

\begin{equation}
    \Pi_{e} = \frac{1}{2}\mathbf{u}_e^{\intercal}\mathbf{K}_e\mathbf{u}_e,
\end{equation}

where $\mathbf{u}_e$ and $\mathbf{K}_e$ are the element displacement
vector and stiffness matrix, respectively.  These values were
log-normalized to match the SIREN input range of $[-1, 1]$:
\begin{equation}
    \tilde{\Pi}_e = 2\cdot\frac{\log(\Pi_e) - \min_{e}(\log(\Pi_e))}{\max_{e}(\log(\Pi_e)) - \min_{e}(\log(\Pi_e))} - 1
\end{equation}
yielding the preprocessed conditioning values
$\{\tilde{\Pi}_e\}_{e=1}^{N}$.

To generate the discrete element densities required for the finite
element solver, we sampled the network at $N$ element-specific inputs.
These inputs were formed by concatenating the coordinates of element
centroids $\x_{e}$ with the corresponding preprocessed strain energy
values $\tilde{\Pi}_{e}$, yielding three-dimensional vectors
$\left[\x_{e}, \tilde{\Pi}_{e} \right]$.

The raw network outputs $\rho_{e}$ were amplified by a factor of
10~\citep{zehnder2021} and processed with two filters. First,  we
applied a linear density filter~\citep{bourdin2001} with a radius of
$\frac{1}{32}$ of the domain size to ensure mesh-independent designs and
prevent checkerboard artifacts. We denote the output of this
transformation with $\bar{\rho}_e$. Second, we applied the shifted
sigmoid filter introduced by~\citet{hoyer2019} to enforce the
volume~\eqref{eq:volume} and box~\eqref{eq:box} constraints:

\begin{equation*}
    \tilde{\rho}_{e} = \frac{1}{1 + e^{-\left(\bar{\rho}_{e} - b\right)}},
\end{equation*}
where $b \in \mathbb{R}$ is a scalar bias computed at each iteration via
bisection to satisfy the volume constraint exactly, i.e., the unique $b$
such that $\sum_{e=1}^{N} \tilde{\rho}_{e} v_e = V^{*}$.
This transformation converts the original constrained optimization problem
into an unconstrained one over the network parameters $\params$. While the original
formulation~\eqref{eq:volume} specifies an inequality constraint, this
transformation enforces the volume constraint with equality. For
compliance minimization, optimal designs use the full material budget,
making these formulations equivalent in practice.

\begin{figure*}[!thb]
    \centering
    \includegraphics[width=160mm]{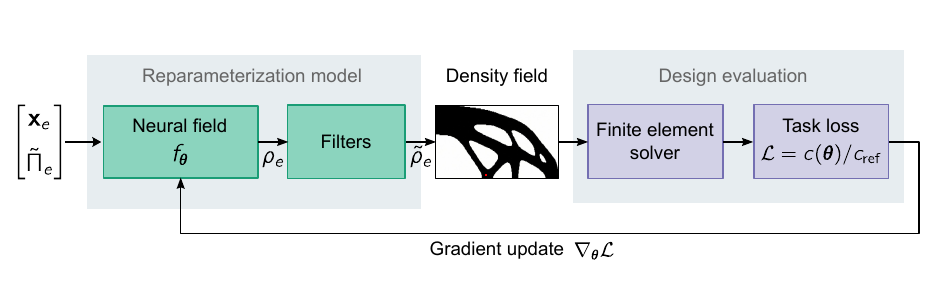}
    \caption{Neural topology optimization (TO) procedure. At each iteration,
        the neural field produces a density field that is filtered and evaluated
        with finite element analysis to compute the compliance loss
        $\mathcal{L}$.}
    \label{fig:framework}
\end{figure*}

The complete neural topology optimization procedure is illustrated
in~\cref{fig:framework}. At each iteration, the network produces a
density field, which is filtered and evaluated with finite element
analysis to compute the compliance $c(\params)$. The task loss is
defined as the compliance normalized by that of a reference design under
the same boundary conditions, $\mathcal{L}=c(\params)/c_{\text{ref}}$,
enabling meaningful comparison across problems with different boundary
conditions and volume fractions. Gradients of the loss with respect to
the parameters of the neural field, $\nabla_{\params}\mathcal{L}$, are
computed with automatic differentiation and used to update the network
parameters via gradient descent. This optimization requires an initial
set of parameters $\params_{0}$, whose choice can significantly
influence both convergence speed and final design quality. Existing
neural TO approaches rely on standard neural network initialization
schemes~\citep{chen2023b} or pretrain the network to represent uniform
density fields~\citep{zhang2021, sanu2024}. These approaches are generic
in that they do not account for the specific problem being solved.

\subsection{Meta-neural TO}
The topology optimization problems considered in this work share the
same design domain, linear elastic material properties, and objective
function, but differ in their boundary conditions and target volume
fractions. We refer to each problem specification as a task $\task$, and
the set of tasks used for the learning process as the meta-training set.
The procedure for pseudorandom tasks used in the experiments is
described in~\cref{sec:task_generation}.

We propose to learn initial parameters $\initparams^{*}$ that enable
fast convergence across a set of related tasks through bilevel
optimization. At the inner level, we consider a subset of tasks, and
perform $k$ optimization steps to minimize their normalized compliance
loss $\mathcal{L}$ (defined in \cref{sec:formulation}). We refer to this
process as adaptation and denote the inner optimizer updates with
$\mathcal{U}_{\text{inner}}^{\task}$:
\begin{align}
    \adaptedparams_{i} \gets \mathcal{U}_{\text{inner}}^{\task}(\initparams;\, \alpha, k), \label{eq:inner}
\end{align}
where $\initparams$ is the initialization shared across tasks and
$\alpha$ is the inner learning rate. At the outer level, we seek initial
parameters $\initparams^{*}$ that minimize the average post-adaptation
loss across all $M$ meta-training tasks:
\begin{align}
    \initparams^{*} = \argmin_{\initparams} \underbrace{\sum_{i=1}^{M}\mathcal{L}(\adaptedparams_{i}(\initparams, \task);\,\task)}_{\eqqcolon\, \mathcal{L}_{\mathrm{meta}}(\initparams)}. \label{eq:outer}
\end{align}
We refer to the outer objective as the meta-loss and denote it
$\mathcal{L}_{\text{meta}}(\initparams)$, dropping the explicit
dependence on the tasks for notational simplicity. We explored two
gradient-based meta-learning algorithms for solving the outer
optimization in~\eqref{eq:outer}: MAML~\citep{finn2017} and
Reptile~\citep{nichol2018}. MAML computes the meta-gradients $\metagrad$
exactly. Since the meta-loss is an implicit function of the inner-loop
gradients, this calculation relied on second-order gradients, which were
prohibitively expensive to compute and store in our context.
Consequently, we adopted Reptile for all experiments in this work.
Reptile approximates the meta-gradients as the difference between the
current initialization and the adapted parameters, averaged over a batch
of $B$ tasks:
\begin{align}
    \metagrad \approx \frac{1}{B}\sum_{i=1}^{B}\left(\initparams - \adaptedparams_{i} \right)\label{eq:reptile_app}
\end{align}
and updates the shared initialization by applying a single step of an
outer optimizer~$\mathcal{U}_{\text{outer}}$:
\begin{align}
    \initparams \gets \mathcal{U}_{\text{outer}}(\initparams;\, \eta) \label{eq:outer_update}
\end{align}
where $\eta$ is the outer learning rate.

\begin{figure}[!thb]
    \centering
    \includegraphics[width=120mm]{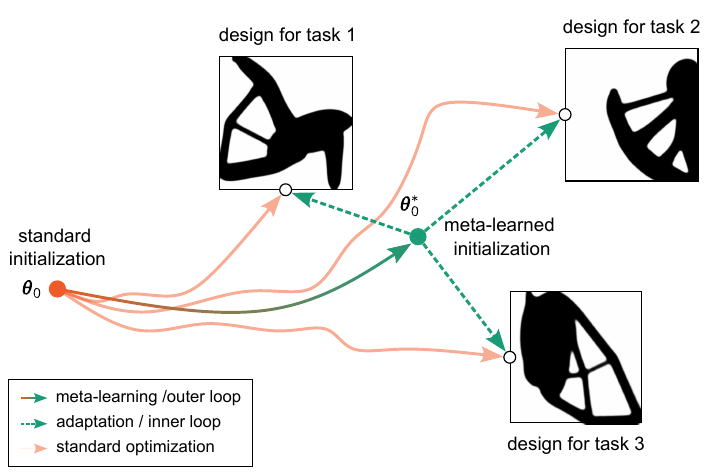}
    \caption{Meta-neural topology optimization (TO) framework. The outer
        loop learns a shared initialization $\initparams^{*}$ from partial
        optimization trajectories across tasks, enabling fast adaptation to
        new design problems.}
    \label{fig:meta_neural_to}
\end{figure}

Unlike supervised approaches that require precomputed datasets of
optimized structures, our method accumulates design experience through
the optimization process itself. The outer loop evaluates initialization
quality based on post-adaptation performance, learning from partial
optimizations rather than from final designs. The meta-learning
procedure is illustrated in~\cref{fig:meta_neural_to} with the
corresponding pseudocode in~\cref{alg:reptile_algo}. 

Each inner-loop iteration in \cref{alg:reptile_algo} (line~8) maps to one cycle of the neural TO procedure shown in \cref{fig:framework}: the neural field produces a density field, which is filtered and evaluated via finite element analysis to compute the compliance loss; the network parameters are then updated based on the gradients. After $k$ such iterations, the adapted parameters $\adaptedparams_{i}$ are
compared to the shared initialization $\initparams$ to form the Reptile
gradient estimate~\eqref{eq:reptile_app}, which drives the outer-loop
update of $\initparams$.

\begin{algorithm}[ht!]
    \caption{Meta-neural topology optimization with Reptile}
    \label{alg:reptile_algo}
    \begin{algorithmic}[1]
        \REQUIRE Meta-training set $\mathcal{S}^{\mathrm{tr}} =
            \left\{\mathcal{T}_{i}\right\}_{i=1}^{M}$
        \REQUIRE Inner learning rate $\alpha$, outer learning rate $\eta$, inner steps $k$, meta-batch size $B$
        \STATE Initialize parameters $\initparams$
        \STATE Initialize outer optimizer (Adam with learning rate $\eta$)
        \FOR{number of outer steps}
        \STATE Sample a batch of tasks $\mathcal{B} =
            \{\mathcal{T}_i\}_{i=1}^B \sim \mathcal{S}^{\mathrm{tr}}$
        \STATE Initialize meta-gradient accumulator $\bm{g} \gets \bm{0}$
        \FOR{each task $\mathcal{T}_i \in \mathcal{B}$}
        \STATE Initialize inner optimizer (Adam with learning rate $\alpha$)
        \STATE Adapt to task: $\adaptedparams_{i} \gets
            \mathcal{U}_{\text{inner}}^{\task}(\initparams;\, \alpha, k)$
        \COMMENT{\eqref{eq:inner}}
        \STATE Accumulate: $\bm{g} \gets \bm{g} +
            (\initparams - \adaptedparams_{i})$
        \ENDFOR
        \STATE Compute meta-gradient: $\metagrad \gets \frac{1}{B} \cdot\bm{g}$
        \COMMENT{\eqref{eq:reptile_app}}
        \STATE Update initialization: $\initparams \gets \mathcal{U}_{\text{outer}}(\initparams;\, \eta)$
        \COMMENT{\eqref{eq:outer_update}}
        \ENDFOR
        \RETURN Learned initialization $\initparams^{*}$
    \end{algorithmic}
\end{algorithm}

\begin{figure}[ht!]
    \begin{minipage}{.44\linewidth}
        \centering
        \includegraphics[width=\linewidth]{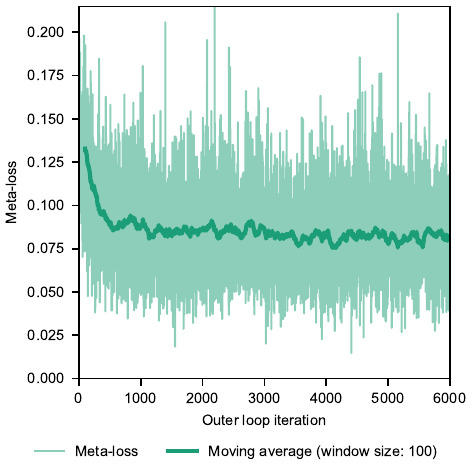}
        \captionof{figure}{Meta-training loss curve. Evolution of the meta-loss
        for the selected hyperparameters ($\omega_0=60.0$, $\eta=10^{-6}$,
        $\alpha=10^{-4}$).}
        \label{fig:training_curve}
    \end{minipage}\hfill
    \begin{minipage}{.52\linewidth}
        \centering
        \small
        \begin{tabular}{lll}
            \toprule
            \textbf{Parameter}      & \textbf{Symbol} & \textbf{Value} \\
            \midrule
            \multicolumn{3}{l}{\textit{Meta-neural TO}}                \\
            Meta-learning algorithm & ---             & Reptile        \\
            Outer optimizer         & ---             & Adam           \\
            Outer learning rate     & $\eta$          & $10^{-6}$      \\
            Outer iterations        & ---             & \num{6000}     \\
            Meta-batch size         & $B$             & 5              \\
            Inner optimizer         & ---             & Adam           \\
            Inner learning rate     & $\alpha$        & $10^{-4}$      \\
            Inner steps             & $k$             & 10             \\
            Frequency parameter     & $\omega_0$      & 60             \\
            \midrule
            \multicolumn{3}{l}{\textit{Neural TO}}                     \\
            Frequency parameter     & $\omega_0$      & 30             \\
            Learning rate           & ---             & $10^{-4}$      \\
            \bottomrule
        \end{tabular}
        \captionof{table}{Hyperparameters for meta-neural TO and the neural TO baseline. Values were selected based on validation performance across all tested configurations.}
        \label{tab:meta_hyperparams}
    \end{minipage}
\end{figure}

We chose the Adam optimizer~\citep{kingma2017} for both the inner and
outer loop updates. In line with previous works in neural
TO~\citep{herrmann2024, sanu2024}, we observed that simple gradient
descent was unable to navigate nonconvex loss landscapes induced by
reparameterization. The inner and outer learning rates, $\alpha$ and
$\eta$, and the frequency parameter $\omega_{0}$ of the Residual SIREN
model were selected through a grid search on a held-out validation set,
as detailed in~\cref{sec:hyperparameter_selection}, with the final
selection summarized in~\cref{tab:meta_hyperparams}. The inner loop
consisted of $k=10$ iterations, striking a balance between computational
efficiency and effective task-specific adaptation. We trained Reptile
for~\num{6000} outer iterations with a meta-batch size of $B=5$, chosen
so that each task in the meta-training dataset was sampled exactly once
per epoch. Extending training beyond this point led to
meta-overfitting, where the initialization specialized on a subset of
tasks at the cost of generalization~\citep{hospedales2022}. The
evolution of meta losses across the meta-training for the selected
hyperparameters is shown in~\cref{fig:training_curve}.

At meta-testing, we optimized $\params$ for each task for a minimum of
10 and a maximum of 200 iterations, starting from the learned
initialization $\initparams^{*}$. We applied a stopping criterion based
on the change in the loss between consecutive iterations:
\begin{equation*}
    |\mathcal{L}(\params_{j}) - \mathcal{L}(\params_{j-1})| < \epsilon(1 + |\mathcal{L}(\params_{j-1})|)
\end{equation*}

where $\mathcal{L}(\params_{j})$ represents the loss at optimization
iteration $j$ and $\epsilon=10^{-5}$. This criterion accounts for the
scale of the objective function and is well-suited for stochastic
optimizers, such as Adam~\citep{martins2021}. Throughout this paper, we
use ``convergence'' to refer to reaching the stopping criterion, as
opposed to the strict fulfillment of the optimality criterion. To ensure
fair comparisons, we apply volume-preserving
thresholding~\citep{sigmund2013, sigmund2022} to the final designs. As
explained in~\cref{sec:thresholding}, thresholding had a
disproportionately positive effect on standard TO designs compared to
neural TO designs.
\subsection{Task generation} \label{sec:task_generation}
To enable meta-learning, we constructed datasets of pseudorandom TO tasks, each
defined by a set of boundary conditions and a target volume fraction.
All tasks were defined on a square design domain discretized by a
structured grid of $64 \times 64$ elements. We generated \num{3000}
meta-training tasks with point loads and supports using the method
proposed in~\citet{sosnovik2017}. Each task was constructed by randomly
sampling the numbers and positions of boundary and load conditions. The
number of support locations and loads per task were sampled from Poisson
distributions (mean of $\mu=1$ for supports, $\mu=2$ for loads), with
minimum counts enforced to ensure well-posed problems (at least two
support locations and one load per task). Supports were placed
preferentially along the domain edges. To enforce a minimum separation
between boundary conditions, the sampling probability was modulated
based on the distance from existing supports. Target volume fractions
were sampled uniformly from $[0.1, 0.5]$, and load directions uniformly
from $[0, 2\pi]$ radians.

\begin{figure*}[!thb]
    \centering
    \begin{minipage}{0.8\linewidth}
        \centering
        \begin{subfigure}[t]{0.45\linewidth}
            \centering
            \subcaption{}
            \includegraphics[width=50mm]{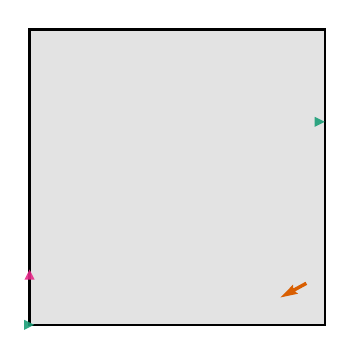}
        \end{subfigure}
        \hfill
        \begin{subfigure}[t]{0.45\linewidth}
            \centering
            \subcaption{}
            \includegraphics[width=50mm]{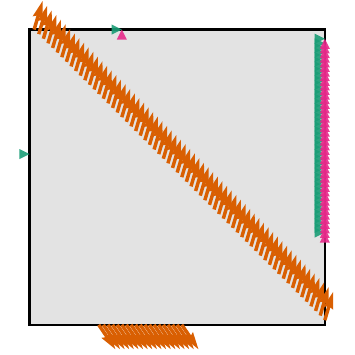}
        \end{subfigure}
    \end{minipage}

    \begin{minipage}{\linewidth}
        \vspace{3mm}
        \centering
        \includegraphics[width=60mm]{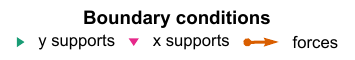}
    \end{minipage}
    \caption{Example boundary conditions. Tasks from the \textbf{(a)}
        in-distribution dataset with point loads and supports, and \textbf{(b)}
        out-of-distribution dataset with distributed boundary conditions along
        line segments.}
    \label{fig:bc_examples}
\end{figure*}

We constructed three test datasets of \num{1000} tasks each to evaluate
different levels of generalization: (i) an in-distribution dataset with
point loads and supports matching the training distribution, (ii) an
out-of-distribution dataset where a subset of boundary conditions were
distributed along randomly generated line segments on the domain
boundary, following~\citet{maze2022}, and (iii) a cross-resolution
dataset with point loads and supports on a $256\times256$
discretization. Further, we generated 100 tasks for hyperparameter
selection using the same generation process as for the meta-training
data. Examples of boundary conditions in in-distribution and
out-of-distribution datasets are shown in \cref{fig:bc_examples}. When
generating the datasets, we performed one finite element analysis per
task to calculate the strain energy field $\tilde{\boldsymbol{\Pi}}$ and
the reference compliance $c_{\text{ref}}$. Conveniently, this enabled us
to filter out tasks with rigid-body modes by identifying singular matrix
errors in the equilibrium solves.

\section{Results}
\begin{figure*}[ht!]
    \begin{minipage}{\linewidth}
        \centering
        \begin{subfigure}[t]{0.3\linewidth}
            \subcaption{}
            \centering
            \includegraphics[width=60mm]{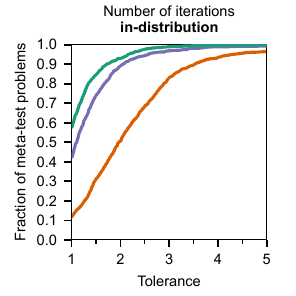}
            \label{fig:id_steps}
        \end{subfigure}
        \hfill
        \begin{subfigure}[t]{0.3\linewidth}
            \subcaption{}
            \centering
            \includegraphics[width=60mm]{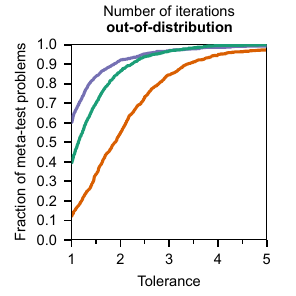}
            \label{fig:ood_steps}
        \end{subfigure}
        \hfill
        \begin{subfigure}[t]{0.3\linewidth}
            \subcaption{}
            \centering
            \includegraphics[width=60mm]{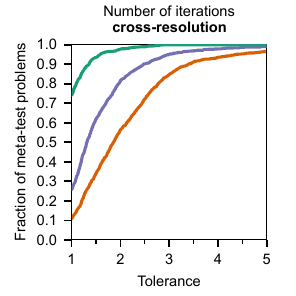}
            \label{fig:upscaled_steps}
        \end{subfigure}

        \begin{subfigure}[t]{0.3\linewidth}
            \subcaption{}
            \centering
            \includegraphics[width=60mm]{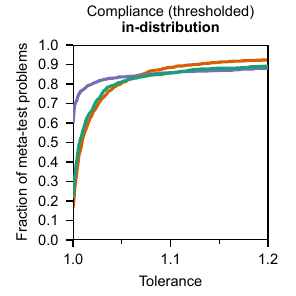}

            \label{fig:id_loss_thr}
        \end{subfigure}
        \hfill
        \begin{subfigure}[t]{0.3\linewidth}
            \subcaption{}
            \centering
            \includegraphics[width=60mm]{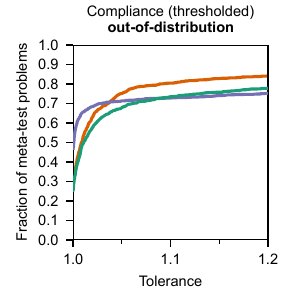}
            \label{fig:ood_loss_thr}
        \end{subfigure}
        \hfill
        \begin{subfigure}[t]{0.3\linewidth}
            \subcaption{}
            \centering
            \includegraphics[width=60mm]{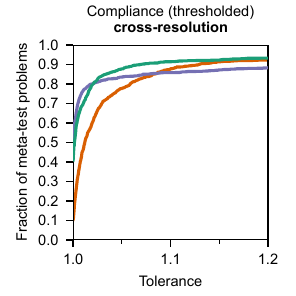}
            \label{fig:upscaled_loss_thr}
        \end{subfigure}
    \end{minipage}
    \hfill
    \begin{minipage}{\linewidth}
        \centering
        \includegraphics[width=100mm]{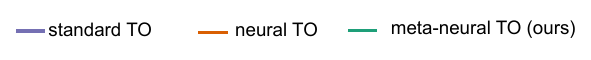}
    \end{minipage}
    \caption{Performance profiles for meta-neural TO and baselines. Top
        row (\textbf{a--c}): number of iterations to convergence. Bottom
        row (\textbf{d--f}): thresholded design compliance. Columns
        correspond to in-distribution, out-of-distribution, and
        cross-resolution experiments, respectively.}
    \label{fig:perfprof}
\end{figure*}

We evaluated our approach on three test datasets, examining \textbf{(i)}
whether meta-learned initializations can consistently accelerate
convergence, \textbf{(ii)} how they affect the final design compliance,
and \textbf{(iii)} how they generalize to tasks outside of the training
distribution. Two additional datasets evaluating robustness to domain
geometry and compound distribution shifts are reported in
\cref{sec:additional_datasets}. We evaluated meta-neural TO against two baselines:
\textbf{(i)} standard density-based TO without neural reparameterization
solved using the globally-convergent version of the MMA algorithm
(GCMMA) with default NLopt settings~\citep{johnson2007} (no
hyperparameter tuning), and \textbf{(ii)} neural TO (as described in
\cref{sec:neural_to}) with a principled SIREN initialization
scheme~\citep{sitzmann2020} and hyperparameters tuned on the validation
set ($\omega_0=30$, the best-performing configuration for standard
neural TO; see~\cref{sec:hyperparameter_selection}). To ensure fair
comparison, we report single runs for meta-neural TO and standard TO, as
their performance is deterministic given the strain energy conditioning
and problem setup, respectively. To present the strongest possible
baseline, we ran neural TO with three different random seeds and
reported the best-case results, i.e., those from the seed that required
the fewest iterations to converge on the in-distribution dataset. The
baselines offer complementary perspectives for assessing our method.
Neural TO helps to isolate the effect of meta-learned initialization,
while standard TO provides context for the practical relevance of our
approach. We compared the methods using performance
profiles~\citep{dolan2004}, as shown in \cref{fig:perfprof}. For each
task, we identified the best-performing method(s) and assessed the
performance of other methods relative to it. The profiles illustrate the
fraction of tasks each method solves within a tolerance of the best
result. For example, a performance ratio of 0.9 at the tolerance of 1.5
indicates that a given method solves 90\% of the test tasks within a
50\% error of the best performer for each task. The profiles reveal how
often an algorithm is the top performer ($\text{tolerance} = 1$) and how
closely it trails the leader (at $\text{tolerance} > 1$). We direct the
readers to~\citet{dolan2004} for methodological details on performance
profiles and~\citet{rojas-labanda2015} for an application in TO.

\subsection{In-distribution}
Meta-neural TO converged in the fewest iterations in 57.6\% of
in-distribution tasks, followed by standard TO at 42.3\% and neural TO
at 11.7\% (\cref{fig:id_steps}). The sum of performance ratios exceeds
100\% here because of ties, primarily on tasks where all methods use the
maximum budget of 200 iterations and therefore tie for the fastest
convergence. This only occurred for 48
tasks, confirming that the 200-iteration budget was sufficient for
performance comparisons. On average, meta-neural TO required 103.65
iterations, standard TO 111.26, and neural TO 149.56 (see
\cref{tab:avg_iterations}). These results indicate that meta-neural TO not only matches the
optimization efficiency of the conventional approach but can surpass it.

\begin{figure}[!ht]
    \centering
    \includegraphics{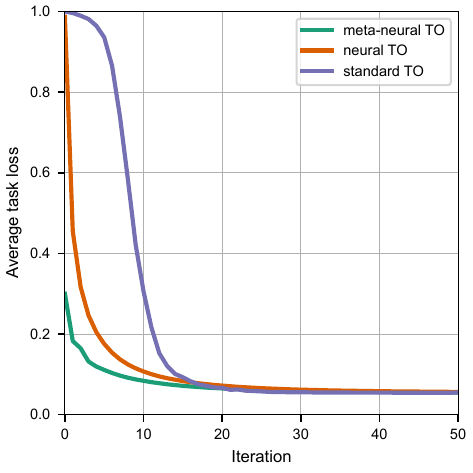}
    \caption{Average convergence curves. Mean normalized compliance loss over the in-distribution test tasks as a function of optimization iteration.}
    \label{fig:avg_loss_id}
\end{figure}

\cref{fig:avg_loss_id} shows how the average loss over the
in-distribution tasks evolves across meta-testing iterations. While both
neural approaches demonstrate rapid initial loss reduction, meta-neural
TO starts at a lower loss value and maintains the lowest loss throughout
the optimization. Standard TO requires a few iterations before
significant loss reduction occurs. The initial gradual convergence is attributed to
the default parameters in GCMMA, which prioritize robustness over
aggressive early optimization steps. Eventually, all three methods reach
low average loss values, confirming that each approach can reliably find
high-quality solutions across the task distribution.

While standard TO found the best thresholded designs in 63.1\% of tasks,
compared to 20.3\% for meta-neural TO and 16.6\% for neural TO
(\cref{fig:id_loss_thr}), all methods performed comparably within a 5\%
tolerance, with performance ratios of 83.9\%, 80.9\%, and 80.4\% for
standard TO, meta-neural TO and neural TO, respectively. Meta-neural TO
frequently produced designs with quality comparable to standard TO. In
fact, the analysis in \cref{sec:thresholding} shows that meta-neural TO
found the best continuous designs in $69.1\%$ of tasks.

\subsection{Out-of-distribution}
Meta-neural TO was fastest on 39.3\% of out-of-distribution tasks,
outperforming neural TO at 12.5\% but trailing standard TO at 59.9\%
(\cref{fig:ood_steps}). Meta-neural TO required an average of 126.57
iterations --- fewer than neural TO (154.67) but more than standard TO
(110.89) (see \cref{tab:avg_iterations}). The quality of thresholded
designs showed a similar trend (\cref{fig:ood_loss_thr}) --- meta-neural
TO found superior designs in 24.6\% of tasks versus standard TO's
47.6\%, with both methods achieving similar success rates at a 5\%
tolerance.

\subsection{Cross-resolution}
\begin{figure}[!tb]
    \centering\includegraphics[width=120mm]{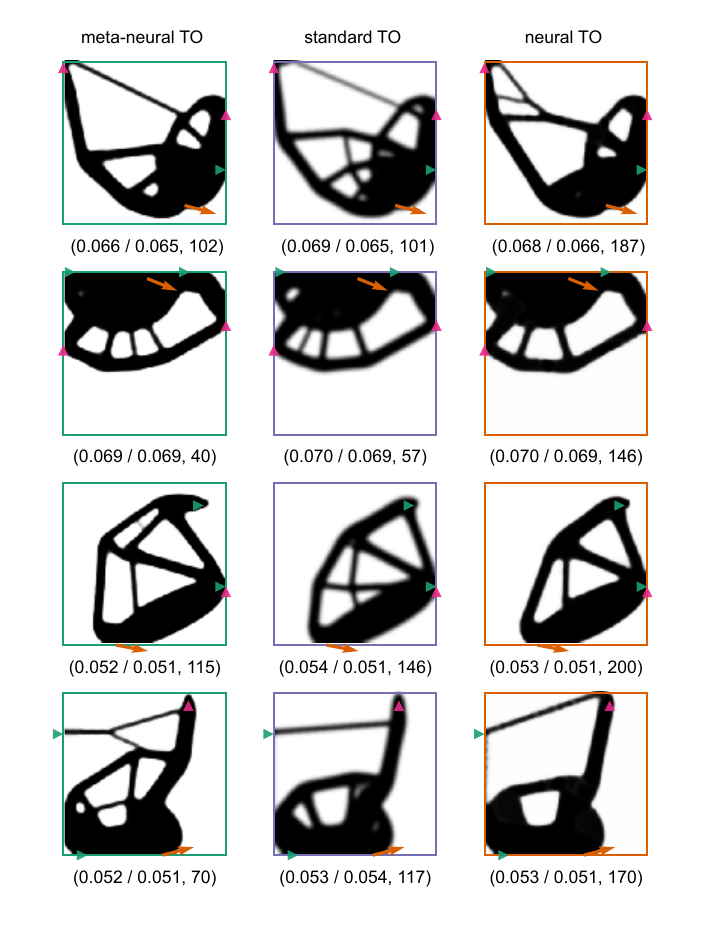}
    \caption{Cross-resolution design examples. Each row shows a test case with optimized structures and performance metrics reported as (normalized compliance before/after thresholding, iterations to convergence).}
    \label{fig:gallery}
\end{figure}

Meta-neural TO excels in cross-resolution experiments, where it was
applied to tasks with finer discretization than used at meta-training.
It required the fewest iterations in 74.1\% of tasks compared to 25.7\%
for standard TO and 10.8\% for neural TO (\cref{fig:upscaled_steps}).
On average, meta-neural TO converged in 100.90 iterations, standard TO
in 130.03, and neural TO in 152.04 (see \cref{tab:avg_iterations}) ---
the learned initialization reduced the average number of iterations by
33.6\% compared to neural TO. \cref{fig:gallery} showcases examples from
our cross-resolution test set, where meta-neural TO consistently
achieves the lowest compliance values while requiring significantly
fewer iterations than both baseline methods.

This experiment highlights the advantageous scaling of neural
representations --- while standard TO must handle 16 times more design
variables, neural TO benefits from a mesh-independent parameterization.
The improved efficiency did not come at the cost of final design quality,
as meta-neural TO and standard TO found competitive final thresholded
designs, winning in 41.0\% and 48.5\% of tasks, respectively
(\cref{fig:upscaled_loss_thr}). At a 5\% tolerance, meta-neural TO
achieved a performance ratio of 87.7\%, surpassing standard TO's 83.6\%.
These results demonstrate an effective generalization of meta-learned
initializations to discretizations finer than those seen during
meta-training.

\begin{table}[htb!]
    \centering
    \small
    \begin{tabular}{lccc}
        \toprule
                       & In-distribution & Out-of-distribution & Cross-resolution \\
        \midrule
        Standard TO    & 111.26          & \textbf{110.89}     & 130.03           \\
        Neural TO      & 149.56          & 154.67              & 152.04           \\
        Meta-neural TO & \textbf{103.65} & 126.57              & \textbf{100.90}  \\
        \bottomrule
    \end{tabular}
    \caption{Average number of iterations to convergence across test datasets. Bold values indicate the lowest average for each scenario.}
    \label{tab:avg_iterations}
\end{table}
\subsection{Representative designs}
\begin{figure*}[ht!]
    \centering
    \includegraphics[width=160mm]{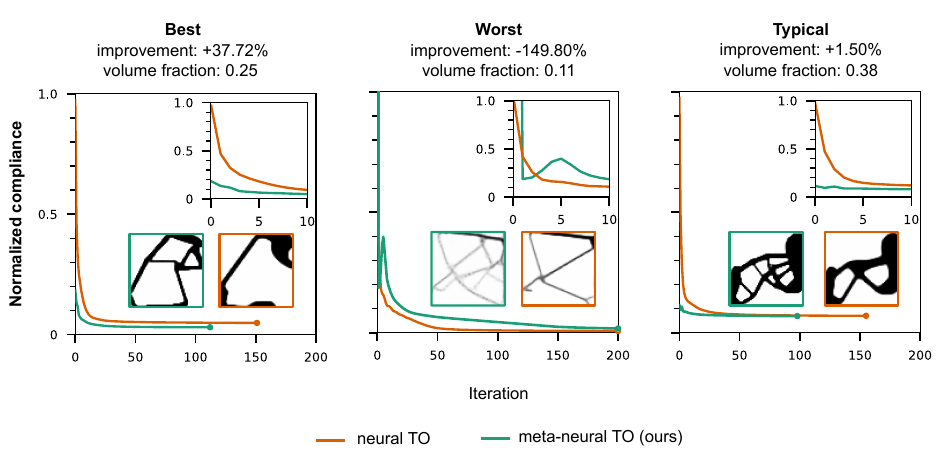}
    \caption{Representative in-distribution tasks. Convergence curves and final designs for tasks with the best, worst, and median improvement in continuous compliance from meta-neural TO over neural TO.}
    \label{fig:id_test_designs}
\end{figure*}

\begin{figure}[!ht]
    \centering
    \includegraphics{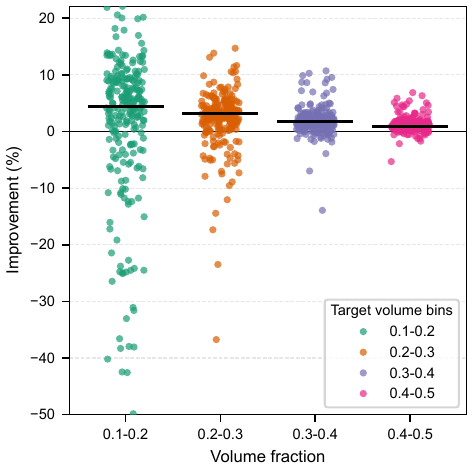}
    \caption{Compliance improvement by volume fraction. Improvement of meta-neural TO over neural TO on in-distribution tasks, grouped by target volume fraction. Black lines indicate bin medians.}
    \label{fig:improvements_vs_volume}
\end{figure}

\cref{fig:id_test_designs} illustrates the convergence curves and final
designs (before thresholding) for a set of representative
in-distribution tasks. These tasks correspond to the best, worst, and
typical (median) improvements in the final design quality of meta-neural
TO compared to neural TO. The best and typical (median) designs achieved
37.72\% and 1.50\% improvements in compliance, respectively. Notably, in
both cases, the meta-learned initialization corresponds to high-quality
initial designs (as indicated by low compliance at the 0th iteration).
The final designs from meta-neural TO exhibited finer structural
features, most likely due to the higher frequency parameter used in the
model ($\omega_0=60$ for meta-neural TO versus $\omega_0=30$ for neural
TO). The worst design showed a drastic 149.8\% increase in compliance.
In this case, the meta-learned initial design underperformed a generic
initialization and biased the optimization towards a poor local minimum.
Notably, neither of the two approaches converged within the
200-iteration budget. As shown in \cref{fig:improvements_vs_volume}, while
meta-learned initializations generally improved performance across
volume fractions (positive median values), tasks with lower target
volumes (0.1-0.2) exhibited higher variability, with both the largest
potential improvements and deterioration in final design quality.

\subsection{Strain energy as a design prior}
\begin{figure*}[ht!]
    \centering
    \includegraphics[width=140mm]{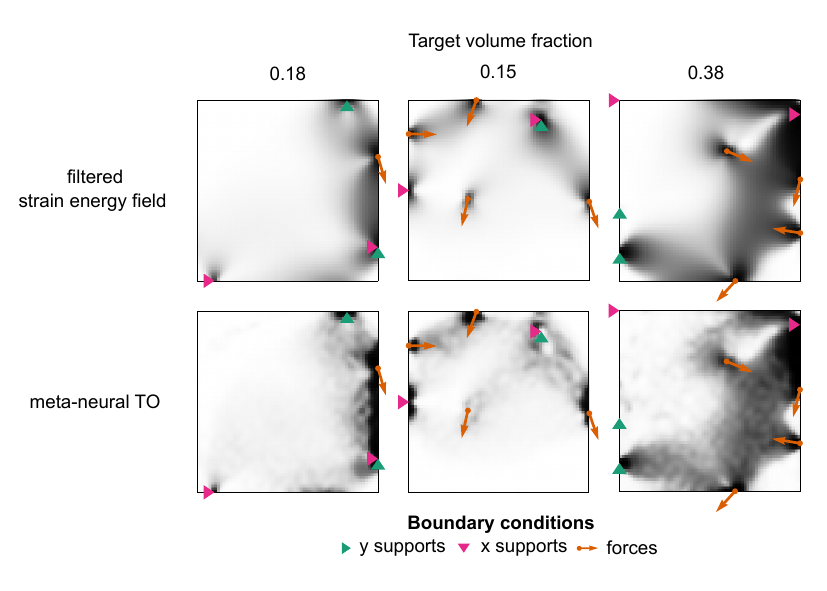}
    \caption{Meta-learned initial designs as strain energy priors. Initial designs from meta-neural TO (bottom) closely mirror the filtered strain energy density fields of uniform density reference designs (top).}
    \label{fig:initial_designs}
\end{figure*}

The analysis of initial designs from meta-neural TO across different
tasks reveals an interesting connection. As shown
in~\cref{fig:initial_designs}, the initial density fields closely mirror
the ones obtained by directly applying the volume and density filters to
the input strain energy fields $\tilde{\boldsymbol{\Pi}}(\x)$. While the
network is conditioned on strain energy fields, it was not explicitly
trained to reproduce them as initial designs. Instead, the meta-learning
process discovered that directly mapping the filtered strain energy
patterns provided an effective initialization strategy across diverse
optimization tasks.

This discovery has a clear physical interpretation. The sensitivity of
the compliance objective with respect to element densities is
proportional to the local strain energy density:
\begin{align}
    \frac{\partial c}{\partial \rho_{e}} = -p\rho_{e}^{p-1}\mathbf{u}_{e}^{\intercal}\mathbf{K}_{e}(\rho_e)\mathbf{u}_{e}
    \label{eq:sensitivity}
\end{align}

Consequently, placing material where strain energy density is high and
removing it where strain energy density is low is a natural strategy for
compliance minimization. This principle underlies several established TO
approaches, including the bubble method~\citep{eschenauer1994} and
bi-directional evolutionary structural optimization~\citep{young1999}.
One might then ask why standard TO does not simply initialize from
strain energy patterns. The reason is that MMA makes conservative local
updates that cannot easily escape an early material layout once the
design approaches a solid/void state~\citep{sigmund2013}. This
limitation does not apply to neural TO, where the continuous
parameterization permits larger structural rearrangements, a property
that the meta-learner exploits. The role of strain energy conditioning
is examined through an ablation in~\cref{sec:conditioning}.

Given that meta-learning converges to strain-energy-based
initializations, a natural question is whether this prior can be encoded
more directly. To verify this finding, we performed an additional
experiment (see \cref{sec:pretraining}), where we pretrained a neural
field to perform an identity mapping on the strain energy input across
training tasks, i.e.,
$f_{\params}(\boldsymbol{x},\tilde{\boldsymbol{\Pi}}(\x)) \approx
    \tilde{\boldsymbol{\Pi}}(\x)$. Surprisingly, this simple approach
outperformed meta-neural TO in both convergence speed and final design
quality. The improvement can be attributed to the directness of the
pretraining objective: while meta-learning must implicitly discover the
relevance of strain energy patterns through thousands of partial
optimizations, pretraining encodes this mapping exactly, producing
initial designs that more faithfully reproduce the filtered strain
energy fields. In effect, pretraining provides the same physical prior
but with less noise from the meta-optimization process.

\subsection{Computational complexity}
Meta-neural TO and neural TO have an identical per-iteration cost at
test time. Both evaluate the same neural field architecture, followed by
finite element analysis, filtering, and Adam optimizer updates.
Consequently, the iteration counts directly reflect the differences in
total computation time. While standard TO does not require an evaluation
of a neural field, it uses a more complex GCMMA optimizer. We evaluated
the wall-clock times across 200 iterations for three classical benchmark
cases: MBB, cantilever, and tensile
beams~\citep{rojas-labanda2015,sigmund2013}, and collected the average
per-iteration costs in \cref{tab:complexity}.

\begin{table}[htb!]
    \centering
    \small
    \begin{tabular}{lccc}
        \toprule
                                                    & Standard TO & Neural TO & Meta-neural TO \\
        \midrule
        Time per iteration ($64 \times 64$) [ms]    & 46.11       & 62.49     & same as
        neural TO                                                                              \\
        Time per iteration ($256 \times 256$) [ms]  & 723.13      & 672.90    & same as
        neural TO                                                                              \\
        Avg.\ iterations (in-dist., $64 \times 64$) & 111.26      & 149.56    & 103.65
        \\
        Meta-training time                          & ---         & ---       & 3h 35min       \\
        \bottomrule
    \end{tabular}
    \caption{Computational complexity comparison. Wall-clock times were
        measured on a single NVIDIA A100 GPU. Per-iteration times are averaged
        over 200 iterations across three benchmark tasks.}
    \label{tab:complexity}
\end{table}
The per-iteration cost of neural TO is approximately $1.36\times$ that
of standard TO on the $64 \times 64$ mesh. However, on the $256 \times
    256$ mesh, neural TO becomes faster ($0.93\times$ the cost of standard
TO). This performance shift occurs because the computational cost of the
GCMMA optimizer update scales less favorably than the neural network
evaluation at higher resolutions. While our implementations are not
fully optimized, these timings serve as a valid relative comparison.

The meta-training represents a one-time investment of 3 hours and 35
minutes on a single NVIDIA A100 GPU. The average throughput during
meta-training ($\approx 43$ ms/iter) is faster than the instrumented
test-time measurement ($\approx 62$ ms/iter) due to the use of
just-in-time compiled loops and reduced logging overhead. This cost is
amortized over all subsequent test problems. Notably, this cost is
equivalent to approximately \num{1500} full optimizations on the $64
    \times 64$ mesh, which is significantly lower than the cost of preparing
datasets for supervised deep learning approaches. For context,
comparable data-driven methods typically rely on generating datasets
ranging from \num{30000} to over \num{2200000} pre-optimized
samples~\citep{nobari2025}.

\section{Discussion}
We introduced meta-neural TO, a framework that accumulates and transfers
design experience across topology optimization problems via
meta-learning. Unlike pretraining approaches~\citep{herrmann2024,
    zhang2021} and direct-design methods~\citep{woldseth2022}, our method
learns from partial optimizations rather than precomputed datasets,
making it readily extensible to different optimizers and
parameterizations. The cross-resolution experiments further show that
this knowledge transfers across mesh discretizations, a property unique
to neural parameterizations. The two additional datasets
of~\cref{sec:additional_datasets} delimit how far this transfer reaches:
a change in domain aspect ratio leaves the iteration advantage intact
(fewest iterations on 63.6\% of tasks), whereas combining distributed
boundary conditions with a finer mesh erodes it (40.9\% versus 56.0\% for
standard TO) --- yet even in that hardest setting meta-neural TO still
yields the best thresholded designs.

A central finding of this work is that strain energy conditioning is
essential for effective knowledge transfer. The meta-learning process
independently converged to strain-energy-based initializations without
being explicitly guided to do so, and the theoretical justification via
the compliance sensitivity~\eqref{eq:sensitivity} confirms why this
strategy is effective. That the same initializations also accelerate
standard TO (\cref{sec:cross-optimizer}) suggests the discovery of
optimizer-agnostic structural priors with potential impact beyond the
neural TO setting.

The method has some practical limitations. Meta-training demands
substantial computational resources, though training on coarse meshes
mitigates this cost. Additionally, generating a diverse task
distribution requires careful consideration. For compliance
minimization, pseudorandom sampling of boundary conditions suffices, but
for more complex physics, the challenge grows considerably. In fracture
mechanics, for instance, one must ensure that generated tasks actually
exhibit crack initiation and propagation rather than trivial elastic
responses---curating such a dataset is itself a nontrivial problem.
Furthermore, the meta-learned initialization can bias the optimizer
towards poor local minima, as shown by the worst-case in-distribution
task, and performance degrades when test problems deviate substantially
from the training distribution.

The practical value of faster convergence is greatest in domains where
families of related problems are solved repeatedly. In aerospace and
automotive engineering, components are routinely optimized under varying
boundary conditions but with similar underlying physics. The topology
optimization of wheel designs~\citep{yoo2025}, where load cases and
symmetry constraints vary across otherwise similar structures, is one
such example. The method is also well-suited for rapid prototyping and
high-resolution problems, where per-iteration cost is high, and
iteration savings translate directly into wall-clock gains. Promising
directions include per-parameter learning rate adaptation via
Meta-SGD~\citep{li2017}, boundary condition similarity metrics to guide
task selection~\citep{lynch2019}, multi-fidelity strategies building on
cross-resolution transfer, and extensions to more complex physics such
as plasticity, fracture, or thermal problems---where initialization is
known to strongly influence which local optimum the solver
reaches~\citep{yan2018}.

\section*{CRediT authorship contribution statement}
\textbf{Igor Kuszczak:} Conceptualization, Methodology, Investigation,
Software, Writing -- original draft. \textbf{Gawe\l{} Ku\'{s}:} Software,
Data curation, Writing -- review \& editing. \textbf{Federico Bosi:}
Supervision, Writing -- review \& editing, Funding acquisition.
\textbf{Miguel A. Bessa:} Conceptualization, Supervision, Writing --
review \& editing, Funding acquisition.

\section*{Declaration of competing interest}
The authors declare that they have no known competing financial
interests or personal relationships that could have appeared to
influence the work reported in this paper.

\section*{Acknowledgements}
This research was supported by the European Union's Horizon 2020
Research and Innovation programme through the Marie Sk\l{}odowska-Curie
grant (Agreement No.\ 956547-LIGHTEN). M.A.B.\ would like to acknowledge
that this effort was undertaken in part with the support from the
Department of the Navy, Office of Naval Research, award number
N00014-23-1-2688. G.K.\ would like to acknowledge that this effort was
undertaken in part with the support from the Department of the Navy,
Office of Naval Research, award number N00014-21-1-2670. I.K.\
acknowledges Suryanarayanan Manoj Sanu and Leon Herrmann for their
valuable feedback on the manuscript. Special thanks to Louis Kirsch for
the discussions that laid the foundation for this work.

\section*{Data availability}
The code used to generate the results presented in this paper is
available at \url{https://github.com/bessagroup/metatopia}. The complete
dataset used in this study is publicly accessible and can be downloaded
from Zenodo (DOI: 10.5281/zenodo.15172934).

\section*{Declaration of generative AI and AI-assisted technologies in the writing process}
During the preparation of this work the authors used Claude (Anthropic) to assist with proofreading and language editing, including grammar correction, clarity improvements, and consistency checks across the manuscript. After using this tool/service, the authors reviewed and edited the content as needed and take full responsibility for the content of the published article.

\bibliographystyle{elsarticle-num-names}
\bibliography{bibliography}

\clearpage
\appendix
\counterwithin{figure}{section}
\counterwithin{table}{section}
\setcounter{figure}{0}
\setcounter{table}{0}

\section{Hyperparameter selection} \label{sec:hyperparameter_selection}
Building on previous studies on meta-learning for neural
fields~\citep{tancik2021, sitzmann2020}, we conducted a focused grid
search over the key model parameters --- inner and outer learning rates,
and frequency parameter $\omega_{0}$. We performed full meta-training
for each hyperparameter combination and evaluated the average normalized
compliance loss after 10 iterations on a held-out validation set
(\cref{tab:reptile-results}).

\begin{table*}[!ht]
    \centering
    \small
    \begin{tabular}{@{}lllccc@{}}
        \toprule
        \multirow{2}{*}{Model}            & \multirow{2}{*}{$\omega_0$} &
        \multirow{2}{*}{$\eta$}           & \multirow{2}{*}{$\alpha$}   &
        \multicolumn{2}{c}{Validation loss}                                                                                                      \\
        \cmidrule(lr){5-6}                &                             &           &           & Meta                                & Standard \\
        \midrule
        \multirow{8}{*}{{Residual SIREN}} & 30.0                        & $10^{-5}$ & $10^{-4}$ &
        0.144                             & \textbf{0.114}                                                                                       \\ 
                                          & 30.0                        & $10^{-6}$ & $10^{-4}$ & \cellcolor{green!20} 0.103          &
        \textbf{0.114}                                                                                                                           \\ 
                                          & 30.0                        & $10^{-5}$ & $10^{-5}$ & \cellcolor{green!20} 0.143          & 0.418
        \\ 
                                          & 30.0                        & $10^{-6}$ & $10^{-5}$ & 0.513                               & 0.418    \\ 
                                          & 60.0                        & $10^{-5}$ & $10^{-4}$ & 0.146                               & 0.128    \\ 
                                          & 60.0                        & $10^{-6}$ & $10^{-4}$ & \cellcolor{green!20} \textbf{0.087}
                                          & 0.128                                                                                                \\ 
                                          & 60.0                        & $10^{-5}$ & $10^{-5}$ & \cellcolor{green!20} 0.283          & 0.487
        \\ 
                                          & 60.0                        & $10^{-6}$ & $10^{-5}$ & \cellcolor{green!20} 0.125          & 0.487
        \\ 
        \bottomrule
    \end{tabular}
    \caption{Validation loss comparison across different hyperparameter configurations. Configurations where meta-neural TO outperforms standard neural TO are highlighted in green. Bold values represent the best-performing configurations.}
    \label{tab:reptile-results}
\end{table*}

As summarized in~\cref{tab:reptile-results}, meta-neural TO outperformed
neural TO with standard initialization in 5 out of 8 presented
configurations, with the lowest validation loss at $\omega_{0} = 60.0$,
outer learning rate of $10^{-6}$, and inner learning rate of $10^{-4}$.
Notably, standard neural TO performed best with $\omega_0=30.0$. This
finding aligns with previous work on coordinate-based neural
representations~\citep{tancik2021, tancik2020}, which shows that
meta-learned models can benefit from higher $\omega_0$ values,
suggesting that learned initialization enables an effective use of
higher-frequency components.

\clearpage
\section{Effect of thresholding on design quality}
\label{sec:thresholding} 
\begin{figure*}[!ht]
    \begin{minipage}{\linewidth}
        \centering
        \begin{subfigure}[t]{0.3\linewidth}
            \subcaption{}
            \centering
            \includegraphics[width=60mm]{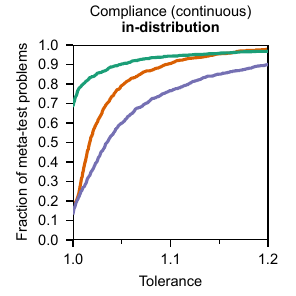}
        \end{subfigure}
        \hfill
        \begin{subfigure}[t]{0.3\linewidth}
            \subcaption{}
            \centering
            \includegraphics[width=60mm]{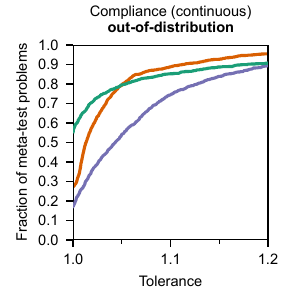}
        \end{subfigure}
        \hfill
        \begin{subfigure}[t]{0.3\linewidth}
            \subcaption{}
            \centering
            \includegraphics[width=60mm]{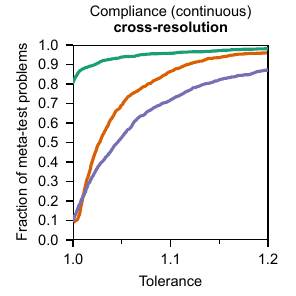}
        \end{subfigure}
        \hfill

    \end{minipage}
    \hfill
    \begin{minipage}{\linewidth}
        \centering
        \includegraphics[width=100mm]{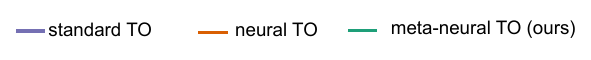}
    \end{minipage}
    \caption{Continuous design compliance. Performance profiles
        comparing continuous (not thresholded) design compliance for
        \textbf{(a)} in-distribution, \textbf{(b)} out-of-distribution, and
        \textbf{(c)} cross-resolution experiments.}
    \label{fig:continuous_perfprof}
\end{figure*}

The performance profiles in
\cref{fig:continuous_perfprof} reveal an interesting pattern where
meta-neural TO finds superior continuous designs for the majority of
tasks across all test datasets. This is most pronounced in the
cross-resolution experiment, where meta-neural TO dominates on 81.0\% of
tasks. Note that these results complement the performance profiles in
\cref{fig:perfprof}. We observed that thresholding had a
disproportionately positive effect on standard TO designs, reducing this
apparent advantage. Here, we report the average improvements in
compliance from thresholding for meta-neural TO and standard TO designs.
For the calculation, we exclude the tasks where thresholding led to a
drastic ($>50\%$) quality degradation. Thresholding improved the
compliance of standard TO designs by 6.63\% (in-distribution) and 5.68\%
(out-of-distribution). For meta-neural TO designs, thresholding improved
compliance by 1.97\% (in-distribution) and 1.74\% (out-of-distribution).
We found that neural TO approaches generally produced designs with fewer
intermediate-density elements, leading to smaller gains from
thresholding. We attribute this effect to the sigmoidal
volume-preserving filter.

\clearpage
\section{Alternative pretraining strategy}
\label{sec:pretraining} 
\begin{figure*}[!ht]
    \begin{minipage}{\linewidth}
        \centering
        \begin{subfigure}[t]{0.3\linewidth}
            \subcaption{}
            \centering
            \includegraphics[width=60mm]{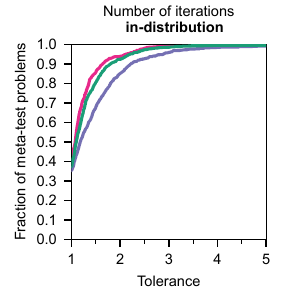}
        \end{subfigure}
        \hfill
        \begin{subfigure}[t]{0.3\linewidth}
            \subcaption{}
            \centering
            \includegraphics[width=60mm]{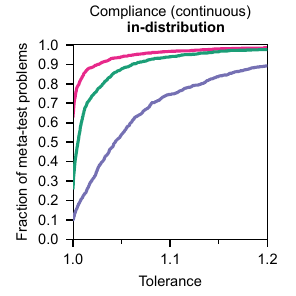}
        \end{subfigure}
        \hfill
        \begin{subfigure}[t]{0.3\linewidth}
            \subcaption{}
            \centering
            \includegraphics[width=60mm]{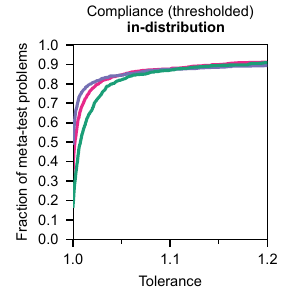}
        \end{subfigure}
        \hfill

    \end{minipage}
    \hfill
    \begin{minipage}{\linewidth}
        \centering
        \includegraphics[width=100mm]{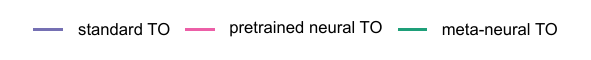}
    \end{minipage}
    \caption{Pretraining versus meta-learning. Performance profiles
        comparing iterations to convergence (\textbf{a}), continuous
        compliance (\textbf{b}), and thresholded compliance (\textbf{c})
        for the pretrained neural TO, meta-neural TO, and standard TO on
        in-distribution tasks.}
    \label{fig:pretrain_perfprof}
\end{figure*}

To better understand the relationship between
strain energy fields and effective initial designs, we compared our
meta-learning approach against a simple pretraining strategy. We trained
the neural field to directly output strain energy density fields (one of
its inputs) using mean squared error loss over 100 epochs (understood as
sweeps through the meta-training dataset). Performance profiles in
\cref{fig:pretrain_perfprof} show that this approach offers moderate
efficiency improvements over the meta-learned initialization, requiring
the fewest iterations in 40.7\% of tasks compared to meta-neural TO's
37.4\%. Further, the approach improves continuous and thresholded
compliance across tasks, narrowing the quality gap between neural and
standard TO.

This approach's strong performance validates a key insight of our
meta-learning framework: the critical importance of strain energy
patterns in generating effective initial designs. Meta-learning
naturally discovered this relationship without explicit supervision,
demonstrating its ability to uncover meaningful physical principles from
the optimization process alone. This discovery mechanism makes our
meta-learning approach particularly promising for applications where the
relevant physical patterns are less well understood or where multiple
fields need to be considered simultaneously.

\clearpage
\section{Meta-neural TO without conditioning}
\label{sec:conditioning} The proposed meta-learning method relies on
strain energy fields to generate task-specific initial designs. Here, we
examine how removing this conditioning mechanism affects the
meta-learning process and resulting performance. Without access to
boundary condition information, the meta-learning objective
fundamentally shifts. Instead of learning to map a physical field into
promising initial designs, the model must discover a universal initial
design that can rapidly adapt to diverse topologies while learning about
problem physics solely through the loss function.

\begin{figure*}[!ht]
    \begin{minipage}{\linewidth}
        \centering
        \begin{subfigure}[t]{0.3\linewidth}
            \subcaption{}
            \centering
            \includegraphics[width=60mm]{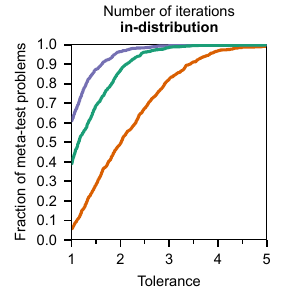}
        \end{subfigure}
        \hfill
        \begin{subfigure}[t]{0.3\linewidth}
            \subcaption{}
            \centering
            \includegraphics[width=60mm]{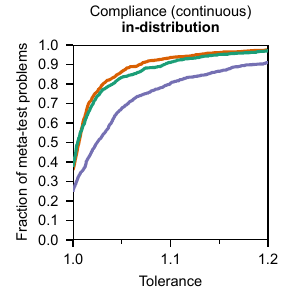}
        \end{subfigure}
        \hfill
        \begin{subfigure}[t]{0.3\linewidth}
            \subcaption{}
            \centering
            \includegraphics[width=60mm]{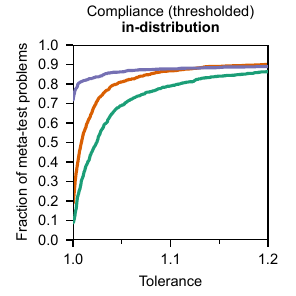}
        \end{subfigure}
        \hfill

    \end{minipage}
    \hfill
    \begin{minipage}{\linewidth}
        \centering
        \includegraphics[width=100mm]{figure_14_legend}
    \end{minipage}
    \caption{Effect of removing strain energy conditioning. Performance
        profiles comparing iterations to convergence (\textbf{a}),
        continuous compliance (\textbf{b}), and thresholded compliance
        (\textbf{c}) for standard TO, neural TO, and meta-neural TO, with
        both neural methods evaluated without strain energy conditioning.}
    \label{fig:nocond_perfprof}
\end{figure*}

\begin{figure*}[!ht]
    \centering
    \includegraphics[width=140mm]{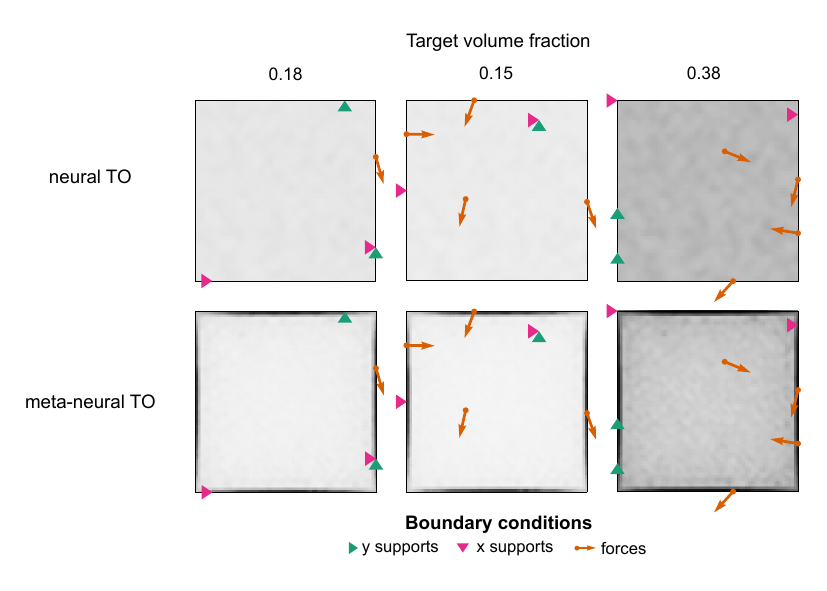}
    \caption{Unconditioned meta-learned initial designs. Without strain energy conditioning, the model concentrates material along domain boundaries, reflecting the statistical distribution of boundary conditions in the training set.}
    \label{fig:init_designs_noncond}
\end{figure*}

Performance profiles in \cref{fig:nocond_perfprof} reveal that removing
strain energy conditioning impacts both the convergence speed and the
final design quality. While meta-learned initialization still
outperforms neural TO with random initialization (both without
conditioning), the improvements are substantially smaller than in the
conditioned case. Notably, the unconditioned meta-learned strategy is
less efficient than standard TO for in-distribution tasks. Analysis of
initial designs in \cref{fig:init_designs_noncond} shows that the
unconditioned model learns to concentrate material along domain
boundaries, mirroring the probability distribution used for sampling
boundary conditions in the meta-training set. This suggests that even
without explicit physical information, meta-learning can discover useful
structural priors from the statistical patterns in the training data.

\clearpage
\section{Can learned initializations benefit standard TO?}
\label{sec:cross-optimizer} 
\begin{figure*}[ht!]
    \begin{minipage}{\linewidth}
        \centering
        \begin{subfigure}[t]{0.3\linewidth}
            \subcaption{}
            \centering
            \includegraphics[width=60mm]{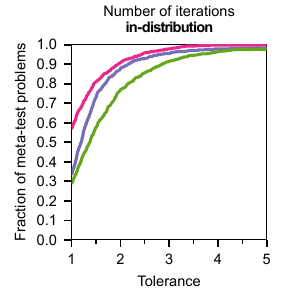}
        \end{subfigure}
        \hfill
        \begin{subfigure}[t]{0.3\linewidth}
            \subcaption{}
            \centering
            \includegraphics[width=60mm]{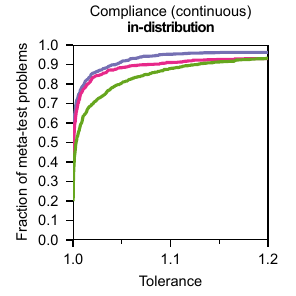}
        \end{subfigure}
        \hfill
        \begin{subfigure}[t]{0.3\linewidth}
            \subcaption{}
            \centering
            \includegraphics[width=60mm]{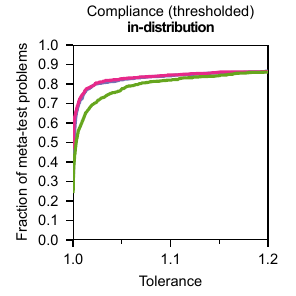}
        \end{subfigure}
        \hfill

    \end{minipage}
    \hfill
    \begin{minipage}{\linewidth}
        \centering
        \includegraphics[width=100mm]{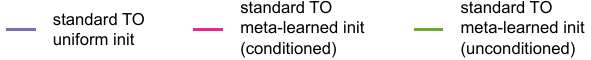}
    \end{minipage}
    \caption{Cross-optimizer transfer. Performance profiles comparing
        standard TO initialized from uniform, conditioned, and
        unconditioned meta-learned designs in terms of iterations to
        convergence (\textbf{a}), continuous compliance (\textbf{b}), and
        thresholded compliance (\textbf{c}).}
    \label{fig:mma_perfprof}
\end{figure*}

Finally, we investigated whether the initial
designs from meta-neural TO can accelerate standard TO. The neural TO
designs satisfy the volume constraint by construction and thus can be
used as initial guesses for the MMA optimizer. The experiments with
initial designs from conditioned and unconditioned meta-neural TO reveal
an interesting dichotomy. As illustrated in \cref{fig:mma_perfprof},
initializing MMA with designs from the strain-energy-conditioned
meta-neural TO improves optimization efficiency. In contrast, initial
designs from the unconditioned model lead to worse performance.

This distinction highlights two key aspects of meta-learning in TO.
First, the meta-learning process inherently considers optimizer dynamics
when searching for improved initializations, making direct transfer
between different optimizers challenging. This explains why
unconditioned initializations, while effective for neural TO, do not
benefit standard TO. Second, when conditioned on strain energy density,
the meta-learned model discovers physically meaningful priors for
material placement that prove beneficial regardless of the optimization
method. These findings underscore both the optimizer-specific nature of
meta-learned initialization strategies and the potential for
physics-informed conditioning to enable cross-optimizer knowledge
transfer.

\clearpage
\section{Evaluation on additional datasets}
\label{sec:additional_datasets}
We introduced two additional datasets to assess how well our method generalizes under domain shift: a
\emph{cross-aspect} dataset featuring a rectangular design domain, and a
\emph{compound-shift} dataset that combines out-of-distribution boundary conditions with higher resolution discretizations from the cross-resolution dataset. Together with the
experiments in the main text, this extended the evaluation to five datasets spanning
a range of aspect ratios, boundary condition types, and mesh resolutions.
All experiments follow the meta-testing protocol outlined in the main text. Each method
runs for a minimum of 10 and a maximum of 200 iterations subject to the relative-change stopping criterion ($\epsilon=10^{-5}$), and all
three methods are compared using performance profiles on the same set of
tasks.

\subsection{Cross-aspect}
The cross-aspect dataset evaluates robustness to changes in domain
geometry. All main-text experiments used a square design domain; here,
we constructed \num{1000} tasks on a rectangular $128 \times 64$ domain
with point loads and supports, following the same generation procedure
as the in-distribution dataset (\cref{sec:task_generation}). Although
the neural-field parameterization naturally accommodates different
domain shapes through its continuous spatial coordinates, it is not
guaranteed that the learned initialization remains effective under this
geometric shift.

\begin{figure*}[ht!]
    \begin{minipage}{\linewidth}
        \centering
        \begin{subfigure}[t]{0.3\linewidth}
            \subcaption{}
            \centering
            \includegraphics[width=60mm]{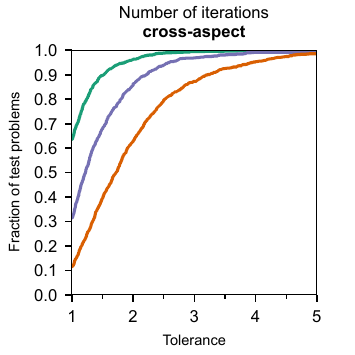}
            \label{fig:ar_steps}
        \end{subfigure}
        \hfill
        \begin{subfigure}[t]{0.3\linewidth}
            \subcaption{}
            \centering
            \includegraphics[width=60mm]{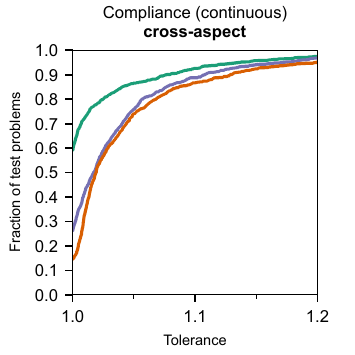}
            \label{fig:ar_loss}
        \end{subfigure}
        \hfill
        \begin{subfigure}[t]{0.3\linewidth}
            \subcaption{}
            \centering
            \includegraphics[width=60mm]{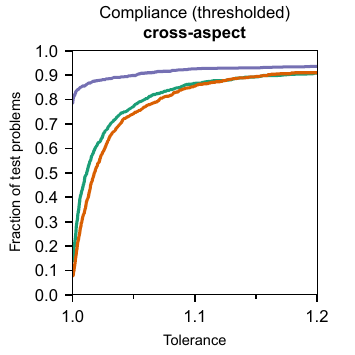}
            \label{fig:ar_bw_loss}
        \end{subfigure}
    \end{minipage}
    \hfill
    \begin{minipage}{\linewidth}
        \centering
        \includegraphics[width=100mm]{figure_17_legend}
    \end{minipage}
    \caption{Performance profiles on the cross-aspect dataset.
        Iterations to convergence (\textbf{a}), continuous compliance
        (\textbf{b}), and thresholded compliance (\textbf{c}).}
    \label{fig:ar_perfprof}
\end{figure*}

Performance profiles for the cross-aspect dataset are shown in
\cref{fig:ar_perfprof}. Meta-neural TO required the fewest iterations
in 63.6\% of tasks, ahead of standard TO at 31.5\% and neural TO at
11.6\% (\cref{fig:ar_steps}), and produced the best continuous designs
in 59.1\% of tasks, ahead of standard TO (26.2\%) and neural TO
(14.7\%) (\cref{fig:ar_loss}). Although the ordering matches the
in-distribution experiment, two differences stand out.

First, meta-neural TO's iteration lead grows relative to
in-distribution (from 57.6\% to 63.6\%) while standard TO's share drops
(from 42.3\% to 31.5\%). As in the cross-resolution experiment,
standard TO must now optimize over twice the number of design
variables, whereas the neural parameterization is mesh-independent.

Second, the thresholded-compliance gap widens in favor of standard TO
(78.5\%, up from 63.1\% in-distribution), with meta-neural TO and
neural TO dropping to 13.8\% and 7.7\% (from 20.3\% and 16.6\%,
respectively, \cref{fig:ar_bw_loss}). Because both neural methods move
in the same direction, we attribute this shift to the neural
parameterization itself rather than to meta-learning. As discussed in
\cref{sec:thresholding}, neural-field designs contain fewer
intermediate-density elements than density-based ones and therefore
benefit less from the volume-preserving filter --- an effect that
widens under the geometric shift.

\subsection{Compound shift}
The compound-shift dataset consists of \num{1000} tasks with
distributed boundary conditions (as in the out-of-distribution dataset)
on a $256 \times 256$ mesh (as in the cross-resolution dataset),
exercising both distribution shifts simultaneously.

\begin{figure*}[ht!]
    \begin{minipage}{\linewidth}
        \centering
        \begin{subfigure}[t]{0.3\linewidth}
            \subcaption{}
            \centering
            \includegraphics[width=60mm]{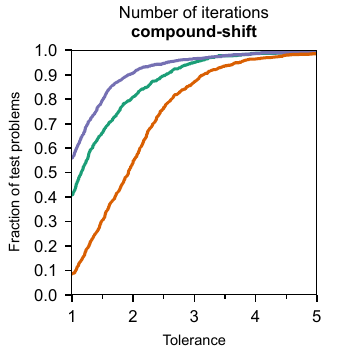}
            \label{fig:cr_ood_steps}
        \end{subfigure}
        \hfill
        \begin{subfigure}[t]{0.3\linewidth}
            \subcaption{}
            \centering
            \includegraphics[width=60mm]{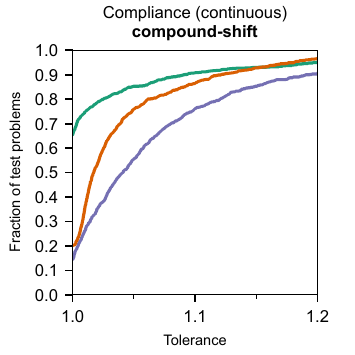}
            \label{fig:cr_ood_loss}
        \end{subfigure}
        \hfill
        \begin{subfigure}[t]{0.3\linewidth}
            \subcaption{}
            \centering
            \includegraphics[width=60mm]{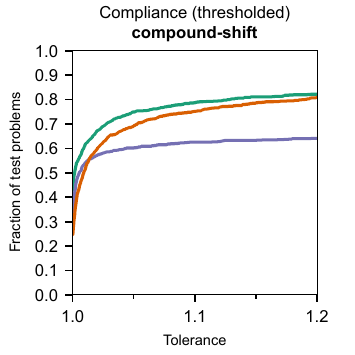}
            \label{fig:cr_ood_bw_loss}
        \end{subfigure}
    \end{minipage}
    \hfill
    \begin{minipage}{\linewidth}
        \centering
        \includegraphics[width=100mm]{figure_17_legend}
    \end{minipage}
    \caption{Performance profiles on the compound-shift dataset.
        Iterations to convergence (\textbf{a}), continuous compliance
        (\textbf{b}), and thresholded compliance (\textbf{c}).}
    \label{fig:cr_ood_perfprof}
\end{figure*}

Performance profiles for the compound-shift dataset are shown in
\cref{fig:cr_ood_perfprof}.
Standard TO required the fewest iterations in 56.0\% of tasks, followed
by meta-neural TO at 40.9\% and neural TO at 8.6\%
(\cref{fig:cr_ood_steps}). Meta-neural TO produced the best continuous
designs in 65.4\% of tasks, ahead of neural TO at 20.1\% and standard
TO at 14.5\% (\cref{fig:cr_ood_loss}), and the best thresholded designs
in 42.7\% of tasks, ahead of standard TO at 32.7\% and neural TO at
24.6\% (\cref{fig:cr_ood_bw_loss}).

The iteration profile resembles the out-of-distribution experiment
(meta-neural TO 39.3\%, standard TO 59.9\%) far more than the
cross-resolution experiment (meta-neural TO 74.1\%). The
distributed-boundary-condition shift dominates iteration behavior, and
adding the resolution shift on top does not recover the speed advantage
that meta-neural TO enjoys on cross-resolution alone.

The thresholded compliance profile is more striking. In both the
out-of-distribution and cross-resolution experiments of the main text,
standard TO won the majority of thresholded comparisons at tolerance $1$
(47.6\% and 48.5\%, respectively), consistent with the broader pattern
that standard TO benefits more from thresholding
(\cref{sec:thresholding}). On the compound-shift dataset, this ordering
reverses, with meta-neural TO winning on 42.7\% of tasks, standard TO
32.7\%, and neural TO 24.6\%. Interestingly, on 9.2\% of tasks,
thresholded standard TO designs perform at least ten times worse than
the best-in-task design, compared to 5.8\% for neural TO and 3.1\% for
meta-neural TO. This severe degradation is linked to a recurring failure
mode where standard TO produces thin, elongated members with diffuse
boundaries that physically disconnect upon thresholding, resulting in
degenerate structures. Meta-neural TO's lead over standard TO is
preserved at a 5\% tolerance (meta-neural TO 74.9\%, neural TO 69.3\%,
standard TO 60.2\%).

\end{document}